# Measurement of dispersive forces between evaporated metal surfaces in the range below 100 nm

P.J. van Zwol, G. Palasantzas[*], M. van de Schootbrugge, and J. Th. M. De Hosson

Department of Applied Physics, Netherlands Institute for Metals Research and Zernike Institute for Advanced Materials, University of Groningen, Nijenborgh 4, 9747 AG Groningen, the Netherlands.

**Abstract**

In this work we describe dispersion force (van der Waals/Casimir) measurements between gold surfaces at separations as close as 12 nm. The force measurements were performed in the plane-sphere configuration by Atomic Force Microscopy at distances ranging between 12 and 200 *nm*. This was accomplished by using evaporated smooth metal surfaces for both sphere and plane, and stiff cantilevers to minimize jump-to-contact problems. Finally, it is shown that below separations of *100 nm* experiment and theory are found in agreement within *10 %* by including in the theory the measured optical properties of gold and contribution from surface roughness.

---

[*]Author to whom correspondence should be addressed: g.palasantzas@rug.nl

Nowadays there is a continuing effort to realize operation of devices at separations between moving parts of a few nanometers under ambient conditions. However, as the proximity between neutral metallic bodies enters the nanometer range, different force fields appear to play significant role. Indeed, these forces lead e.g., to stiction problems, and in more general affect the actuation characteristics of micro/nano electro mechanical systems. At separations below 100 nm, the Casimir force is very strong and becomes comparable to electrostatic forces corresponding to voltages in the range *0.1-1V* [1-3]. On the other hand, from the fundamental point of view, precise measurements of forces at the smallest possible range has attracted considerable interest in a search for new hypothetical force fields [4-7].

Although measurements of the Casimir force at small separations were reported before, these measurements either were limited to separations above *20 nm* [8] (where also problems due to claimed surface deformation were encountered [9]), or they were limited in accuracy [10, 11]. Precision measurements of the Casimir force with claimed accuracy as high as 1% - 0.5 % [7] were performed at separations above *60 nm*, but they remain questionable at separations between *60-100 nm*. This is in particular the case because several key parameters such as the optical properties of real films were not taken into account cautiously , which may well lead to errors above 5 % [12, 13]. Furthermore, errors in the force due to the uncertainty in the separation distance were neglected [14]. The latter effect is inherently unknown with high precision due to surface roughness.

Therefore, we report here a measurement of the Casimir/van der Waals force at separations down to *12 nm* in between surfaces commonly obtained by deposition of metal films*,* and discuss the intrinsic problems with respect to the possible accuracy of such measurement.

Here we will briefly outline the measurement procedure. For details reference is made to [7, 8]. The Casimir force is measured, using the PicoForce AFM (http://www.veeco.com/), between a sphere with a diameter of *100 μm* and an RMS surface roughness *1.2 nm* (attached on a gold coated *240 μm* long cantilever with stiffness *4 N/m*), and a silicon plate. Both sphere and plate are coated with 100 nm Au within the same vacuum evaporator. After Au deposition, the RMS roughness of sphere and plate were measured by AFM to be *1.8* and *1.3 (±0.2) nm* respectively. The optical properties of the Au film on the plate were measured with an ellipsometer in the wavelength range *137 nm – 33 μm* [8]. Fitting the optical data in the infrared range, the Drude parameters $w_p=7.9\pm0.2 eV$ and $w_t=0.048\pm0.005\ eV$ were obtained. This is performed since the optical response and the resulting finite conductivity corrections, for the separations considered here, are very large [12,13]. The roughness spectrum of the films is also included into the theoretical calculations [8, 15].

The calibration of the deflection sensitivity (*m*), cantilever stiffness (*k*), and contact potential ($V_0$) was done in the exact same way as described in our previous work [8] and is based on [7]. Measurement of *m* is done with the plate and sphere in contact while moving the piezo. Electrostatic fitting of *k* and $V_0$ is done within the range of *1 to 4 μm* with voltages in the range ± *(3-4.5) V*. Although we were not able to determine the contact separation due to roughness ($d_0$) electrostatically for the stiff cantilevers employed here, it was derived from the top-to-bottom roughness of sphere and plate (from multiple scans at different places of both surfaces) added and divided by two. This procedure gave for $d_0=7\pm1\ nm$ (Fig. 1). From variations of $d_0$ from location to location on the plane, the error obtained in $d_0$ (derived from a set of roughness scans) is *1 nm* leading therefore to a *~28%* relative error in the force at the smallest separations. Since experimentally the force varies as $F \sim 1/d^2$, i.e. at short separations considered here, the estimate of the relative error due to variation in separation is given by $\Delta F/F \approx 2\Delta d/d$.

The error found for the spring constant $k$ was *4%*, and for the deflection sensitivity *m* was 3%. These two factors lead to errors of 4-10% in the force. The electrostatically obtained contact potential is *10±10 mV* and amounts for up to 10% error for the separations considered here. This is a negligible contribution below *100 nm*. Notably the nonlinearity in our AFM piezo is *0.1%*, and it can be neglected. Noise in the piezo is less than *0.5 nm* and it will average out by repeating the measurement. The sphere diameter is measured with *2 %* precision using the scanning electron microscope (SEM) [16]. All errors together in the calibration lead to an overall error of about 5-35 % in the force measurement as shown in Fig. 2 concluding that the parameters $d_0$ and *m* are the main reasons for the variation seen in Fig 2.

After calibration, the Casimir force is measured and averaged using 40 force curves. The whole procedure, both calibration and measurement as described above is repeated 20 times at different locations on the plane. Thus, we have averaged 800 curves to get the averaged force shown in Fig. 3. If we follow this procedure the uncertainty in $d_0$ and *m* averages out, although we cannot rule out any systematic error in the separation distance *d* due to deformation of the highest surface peaks. This averaging also reduces the error due to thermal noise. Indeed, due to thermal noise the lowest force to resolve in a single curve is $F_{min} \approx (K_B T k)^{1/2}$, yielding for k=4 N/m $F_{min} \approx 100\ pN$. Finally the whole measurement was repeated for a second sphere with the same RMS roughness (see Fig. 3). We should also note that the presence of a linear signal , being not exactly linear over separations in the order of microns, reported for soft cantilevers due to backscattering of light from the surface into the photodiode [7,8], was not observed for our stiff cantilevers. It is likely that soft cantilevers ($k<0.1$ N/m) bend due to stress from the deposited Au, eventually becoming more parallel to the surface, while stiff cantilevers ($k>1$ N/m) appear to be resilient to stress induced bending. Since uncoated cantilevers do not show this signal, it supports our claim.

Fig. 3 shows the averaged force and the relative difference between theory-experiment ($F_{theory}$-$F_{exp}$)/$F_{theory}$. In theory $F_{theory}=(2\pi R/A)E_{pp,rough}$ for plane-sphere geometry, with $E_{pp,rough} = E_{ppflat}+\delta E_{pp,rough}$ the Casimir energy for parallel plates. The Lifshitz's theory yields $E_{ppflat} = -\hbar A \sum_P \int [d^2k/4\pi^2]\int_0^\infty [d\Phi/2\pi]\ln[1-r^p(k,\Phi)^2 e^{-2\kappa d}]$ with A the plane area, r($\Phi$) the reflection coefficient, $\Phi$ the imaginary frequency of the electromagnetic wave, and p the index denoting the transverse electric and magnetic modes. The roughness correction is given by $\delta E_{pp,rough} = \int [d^2k/4\pi^2]G(k)\sigma(k)$ where σ(k) is the roughness power spectrum, and G(k) the response function derived in [15]. For self-affine roughness (typical to occur for non-equilibrium film growth as the Au films here) [17], we consider for σ(k) of the analytic model presented in [18]. The roughness amplitude *w*, the lateral correlation *ξ,* and the roughness exponent *H* are determined by measurement of the height correlation function $H(r) = <[h(r)-h(0)]^2>$ [17]. The second order roughness corrections [15] are *15%* at *12 nm*, and therefore higher order roughness corrections are small in this case. In making the comparison between theory to experiment, we are not fitting the Casimir force. Theory and experiment are in agreement within the *5-10%* level up to *100 nm*. If we fit the Lifshitz theory to our experimental data to obtain $d_0$ we find values of $d_0$ within half a nanometer of our estimated value of $d_0$ for both spheres, from the roughness scans confirming the estimated accuracy of our measurement. At the largest separations the relative error in the force due to thermal noise reaches 100% at 200 nm. Below *12 nm* the cantilever jumps to contact mainly, however, not due to the strong Casimir force but due to attractive capillary forces from the water layer present on the surface under ambient conditions [19].

Although surfaces can be brought into even closer proximity with stiff cantilevers surface roughness and the water layer/capillary condensation will work against this. The water layer on both surfaces is typically a few nm, and even the smoothest evaporated surfaces still have a few nm top to bottom roughness [9], while in addition coatings on the polymer spheres cannot be annealed. Strong capillary forces will also complicate the measurement if surfaces are even smoother (stiction) [16]. In practice all these considerations set a lower limit on the separation between two evaporated metal surfaces, in air, of approximately *10 nm* in which a (measurement) device works properly.

Concluding, we have illustrated the measurement of the Casimir/ van der Waals force under ambient conditions in the range *12-200 nm* using a commercial AFM. Below *100 nm*, the result is in good agreement at the *10 %* level with theory including contribution from measured optical properties and surface roughness of gold coatings. This was made possible by using stiff cantilevers to reduce jump to contact problems. Using more advanced metal deposition techniques such as atomic layer deposition (also for metal coating other than Au, which are used in NEMS/MEMS) [20] further possibilities arise to extend these measurements below *10 nm,* and therefore to further extend our knowledge with respect to force effects in the operation of NEMS/MEMS (e.g., switches).

**Acknowledgements:** The research was carried out under project number MC3.05242 in the framework of the Strategic Research programme of the Netherlands Institute for Metals Research (NIMR). Financial support from the NIMR is gratefully acknowledged.

**References**

[1] H. B. G. Casimir, Proc. K. Ned. Akad. Wet. 60, 793 (1948)

[2] A. Cleland, Foundations of Nanomechanics (Springer, New York, 2003)

[3] H. B. Chan, V. A. Aksyuk, R. N. Kleiman, D. J. Bishop, F. Capasso, Science 291, 1941 (2001)

[4] R. L. Jaffe, Phys. Rev. D 72 (2005), 021301(R)

[5] S. K. Lamoreaux, Phys. Rev. Lett, 78, 5 (1997); S. K. Lamoreaux, Phys. Today, Feb. 2007; S.K. Lamoreaux Phys. Rev. Lett. 83, 3340 (1999).

[6] R. Onofrio, New J. Phys. 8, 237 (2006)

[7] B. W. Harris, F. Chen, U. Mohideen, Phys. Rev. A. 62, 052109 (2000); R. S. Decca, D. López, E. Fischbach, G. L. Klimchitskaya, D. E. Krause, and V. M. Mostepanenko, Phys. Rev. D 75, 077101 (2007); F. Chen, G. L. Klimchitskaya, U. Mohideen, and V. M. Mostepanenko, Phys. Rev. A 69, 022117 (2004); F. Chen and U. Mohideen, G. L. Klimchitskaya, V. M. Mostepanenko, Phys. Rev. A 74, 022103 (2006); F. Chen, U. Mohideen , G.L. Klimchitskaya, V.M. Mostepanenko, Phys. Rev. Lett. 97, 170402 (2006); R. S. Decca, D. López, E. Fischbach, G. L. Klimchitskaya, D. E. Krause, and V. M. Mostepanenko, Annals Phys. 318, 37-80 (2005); R. Decca, E. Fischbach, G. L. Klimchitskaya, D. E. Krause, D. Ló´pez, and V. M. Mostepanenko, Phys. Rev. D 68, 116003 (2003). R.S. Decca[1], D. López, E.Fischbach, G.L. Klimchitskaya, D.E. Krause, and V.M. Mostepanenko, Eur. Phys. J. C 51, 963 (2007). Note that in none of these works direct measurement of the optical properties by ellipsometry was performed. In contrast, either tabulated handbook data were used or limited accuracy conductivity measurements to determine the plasma frequency $w_p$ together with tabulated data for the relaxation $w_t$ in the Drude model were employed.

[8] P.J. van Zwol, G. Palasantzas, J. T. M. De Hosson, To appear in Phys. Rev. B (2008)/ arXiv:0712.1893v1.

[9] T. Ederth, Phys. Rev. A, 62, 062104 (2000)

[10] D. Tabor and R.H. S. Winterton, Proc. R. Soc. Lond. A 312, 435 (1969)

[11] Israelachvili, J. & Tabor, D. Proc. R. Soc. Lond. A 331, 19 (1972).

[12] I. Pirozhenko, A. Lambrecht, V. B. Svetovoy, New J. Phys. 8, 238 (2006); S. L. Lamoreaux, Phys. Rev. A 59, 3149 (1999).

[13] P.J. van Zwol, G. Palasantzas, J. T. M. De Hosson, Appl. Phys. Lett 91 144108 (2007)

[14] D. Iannuzzi, M. Lisanti, F. Capasso, PNAS 101, 4019 (2004)

[15] P. A. Maia Neto, A. Lambrecht and S. Reynaud, Phys. Rev A 72, 012115 (2005)

[16] P.J. van Zwol, G. Palasantzas, J. T. M. De Hosson, Appl. Phys. Lett 91, 101905 (2007); Better accuracy is impossible since SEMS are usually calibrated by such spheres, for which the manufacturer states a few percent deviation in diameter).

[17] J. Krim and G. Palasantzas, Int. J. of Mod. Phys. B 9, 599 (1995); P. Meakin Phys. Rep. 235, 1991 (1994). <...> in $H(r)$ denotes ensemble average over multiple surface scans. For self-affine roughness σ(k) scales as $\sigma(k) \propto k^{-2-2H}$ if $k\xi \gg 1$, and $\sigma(k) \propto const$ if $k\xi \ll 1$. The roughness model we used reads of the form $\sigma(k) = (AHw^2\xi^2)/(1+k^2\xi^2)^{1+H}$ with $A = 2/[1-(1+k_c^2\xi^2)^{-H}]$ and $k_c$ (~1 nm$^{-1}$) a lower roughness cutoff.

[18] G. Palasantzas, Phys. Rev. B 48, 14472 (1993); 49, 5785 (1994); G. Palasantzas and J. Krim, Phys. Rev. Lett. 73, 3564 (1994); G. Palasantzas, Phys.Rev.E. 56, 1254 (1997); Y. -P. Zhao, G. Palasantzas, G. -C. Wang, and J. Th. M. De Hosson, Phys. Rev. B 60, 1216 (1999).

[19] This is because the vdW/Casimir force only could allow more bending of the cantilever to approach down to approximately 9 nm, while jump-to-contact occurs at about 12 nm that can only be explained by the presence of an attractive capillary force.

[20] S. Lim, A. Rahtu, R.G. Gordon, Nature. Materials, 2, 749 (2003)

**Figure captions**

**Figure 1:** Roughness scans and height profiles of plate (left) and sphere (right). The average of the separation upon contact is taken from half the top to bottom roughness of plate and sphere added found from multiple scans.

**Figure 2:** Experimental data of 20 independent measurements compared to theory (solid line). For the roughness parameters we used $w_{sphere}=1.8$ *nm*, $w_{plane}=1.3$ *nm*, $\xi_{sphere,plane}= \sim 20$ *nm*, $H_{sphere, plane}=0.9$.

**Figure 3:** Average of 20 independent measurements shown for 2 different spheres (left) compared to theory (dashed line). The relative error, $(F_{theory}-F_{exp})/F_{theory}$, is shown on the right for the 2 spheres (squares and circles).

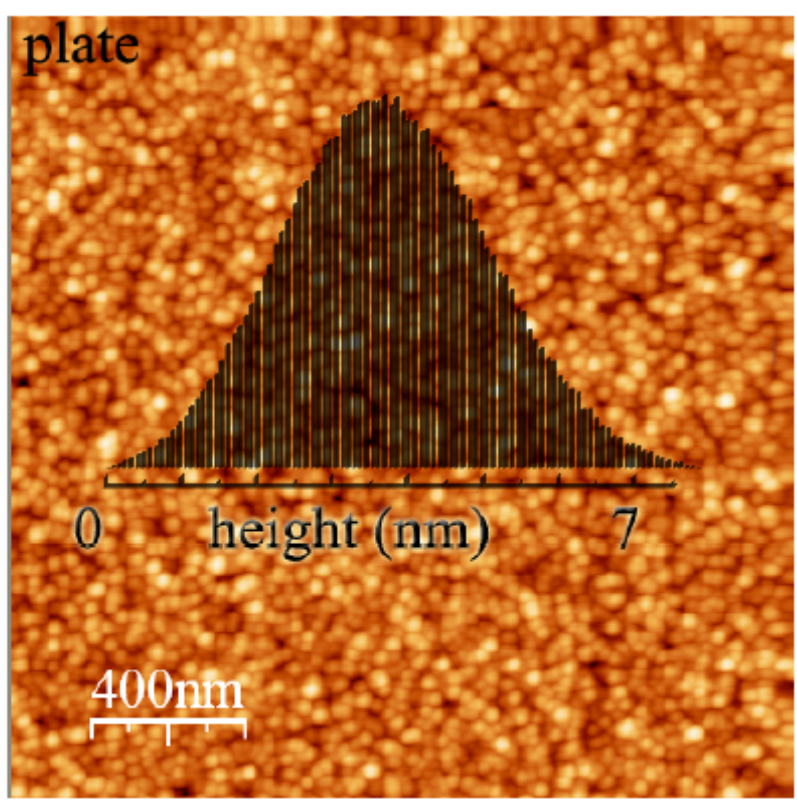 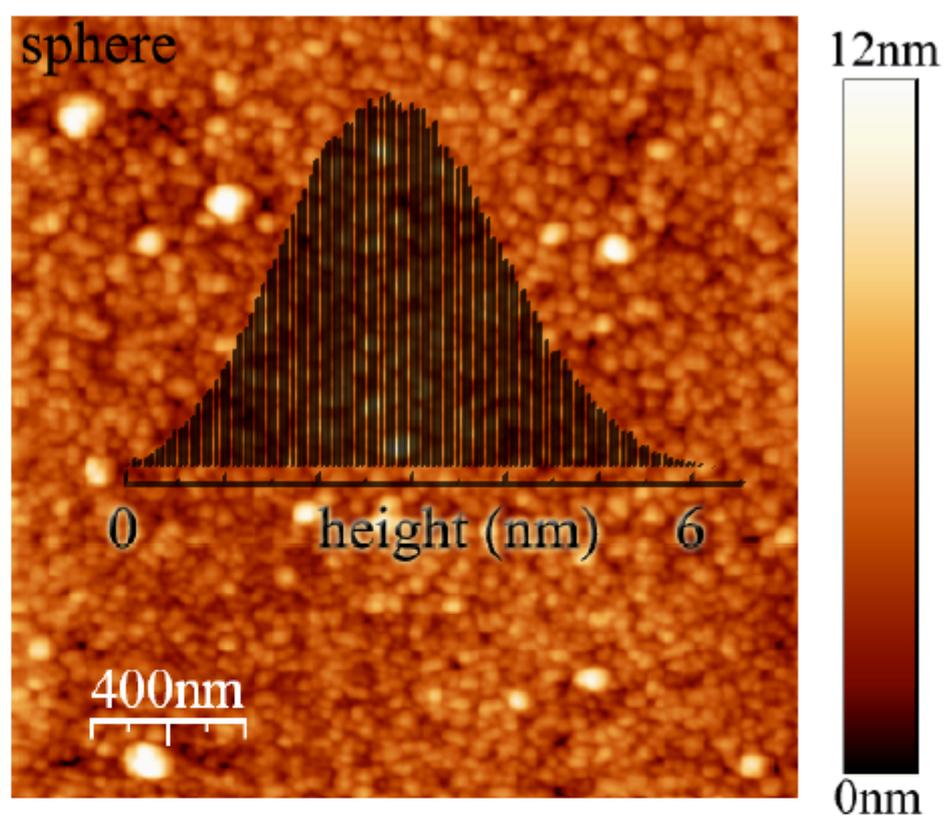

**FIGURE 1**

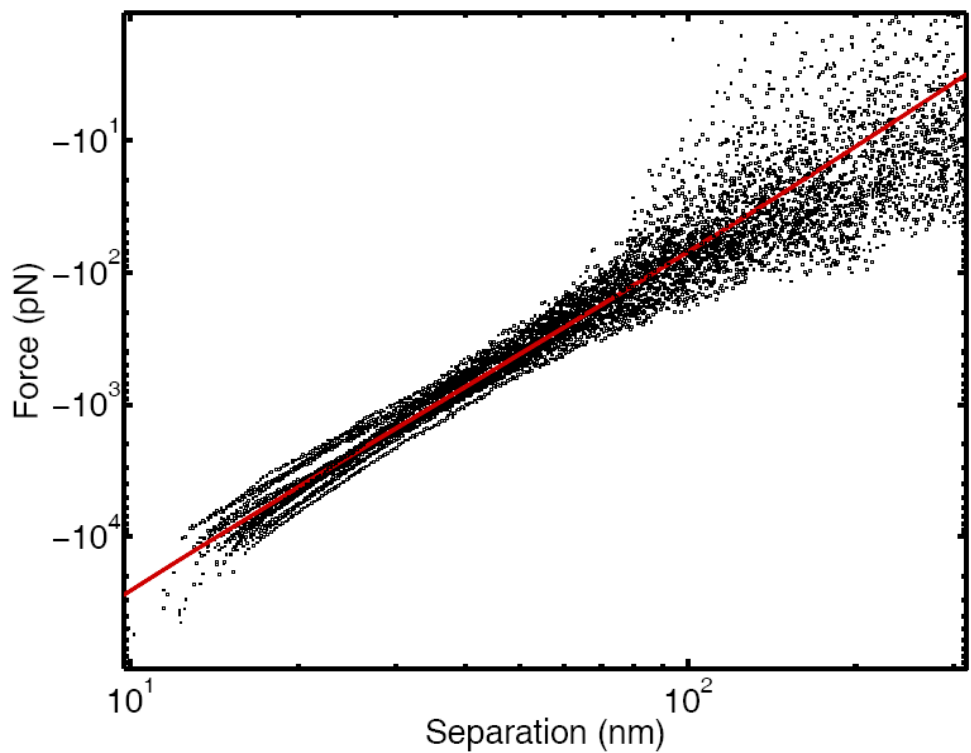

**FIGURE 2**

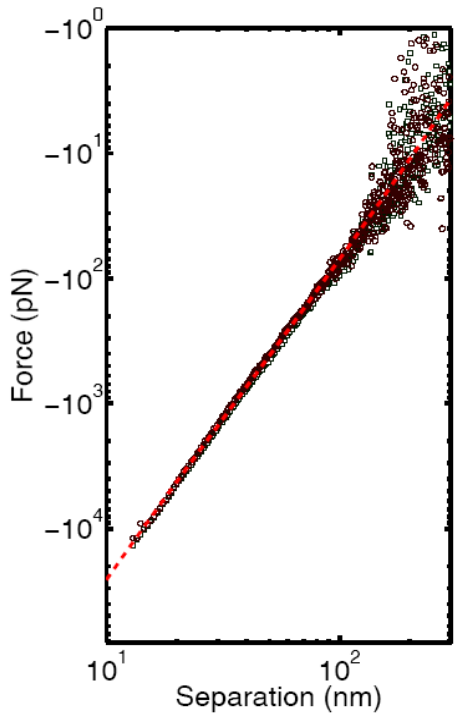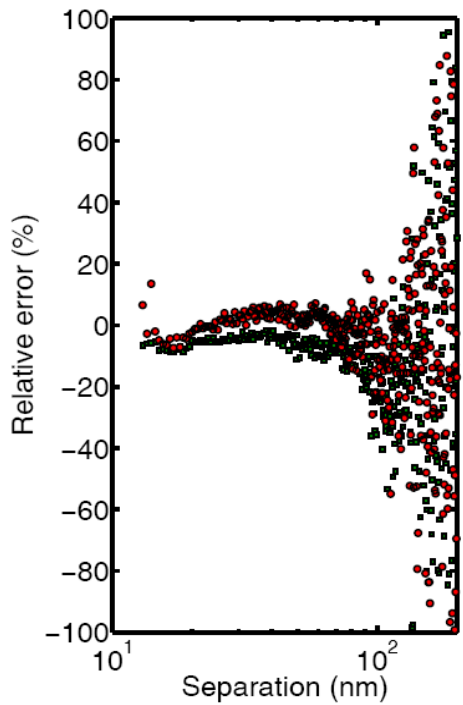

**FIGURE 3**